\documentclass[aps,prl,twocolumn,superscriptaddress]{revtex4-2}
\usepackage{amssymb}
\usepackage{booktabs}
\usepackage{physics}
\usepackage{graphicx}
\usepackage{amsmath}
\usepackage{amsthm}
\usepackage{amsfonts}
\usepackage[T1]{fontenc}
\usepackage{tikz}
\usepackage{array}
\usepackage{multirow}
\usepackage{color}
\usepackage{esint}
\usepackage{bm}
\usepackage{color}
\definecolor{LinkColor}{rgb}{0.75,0.0,0.2}
\usepackage{bbm}
\usepackage{hyperref}
\hypersetup{
	pdfauthor={good guys},
	pdftitle={good title},
	colorlinks=true,
	citecolor=LinkColor,
	linkcolor=LinkColor,
	urlcolor=LinkColor,
}
\usepackage{babel}
\usepackage{titlesec}

\begin{document}
\title{Tunable discrete quasi-time crystal from a single drive}
	
\author{Xu Feng}
\affiliation{Beijing National Laboratory for Condensed Matter Physics, Institute
of Physics, Chinese Academy of Sciences, Beijing 100190, China}
\affiliation{School of Physical Sciences, University of Chinese Academy of Sciences,
Beijing 100049, China }

\author{Shuo Liu}
% \email{sl6097@princeton.edu}
\affiliation{Department of Physics, Princeton University, Princeton, New Jersey 08544, USA}
\affiliation{Institute for Advanced Study, Tsinghua University, Beijing 100084, China}

\author{Shu Chen}
\email{schen@iphy.ac.cn }
\affiliation{Beijing National Laboratory for Condensed Matter Physics,
Institute of Physics, Chinese Academy of Sciences, Beijing 100190, China}
\affiliation{School of Physical Sciences, University of Chinese Academy of Sciences,
Beijing 100049, China }

\author{Shi-Xin Zhang}
\email{shixinzhang@iphy.ac.cn}
\affiliation{Beijing National Laboratory for Condensed Matter Physics, Institute
of Physics, Chinese Academy of Sciences, Beijing 100190, China}

\date{\today}
 
\begin{abstract}
The search for exotic temporal orders in quantum matter, such as discrete quasi-time crystals (DQTCs), has become an important theme in nonequilibrium physics. However, realizing these phases has so far required complex protocols, such as drives with multiple incommensurate frequencies.
Here, we present a significantly simpler mechanism: the emergence of DQTCs in a dissipative collective spin system subjected to only a single periodic drive. Remarkably, the characteristic frequencies of this novel phase are not fixed but can be continuously tuned by varying the strength of the drive. Even more strikingly, this tunability is punctuated by Arnold tongues, within which the response main frequency locks to rational fractions of the drive.
Our model further provides a unified framework that also encompasses stationary, discrete time crystals and chaotic phases. This discovery simplifies the requirements for generating complex temporal orders and opens a viable route towards the experimental control and manipulation of quasi-time crystalline matter.
\end{abstract}

\maketitle
\textit{Introduction.—}
Time-translation symmetry is one of the most fundamental invariances of physics, yet its spontaneous breaking has recently emerged as a defining feature of nonequilibrium quantum matter.  
Motivated by Wilczek’s original proposal of continuous time crystals (CTCs) in closed Hamiltonian systems \cite{WilczekTC,TCsTrappedIons}, the possibility of realizing phases with persistent temporal order was initially met with strong skepticism: a series of no-go theorems demonstrated that continuous time-translation symmetry breaking cannot occur in equilibrium short-range systems \cite{NoGoTheorem1,NoGoTheorem2,NoGoTheorem3,CommentTCsTrappedIons}.
These constraints shifted attention toward nonequilibrium settings, where the symmetry-breaking mechanism can circumvent the limitations of equilibrium.
The subsequent discovery of time crystals, characterized by long-range temporal order, revealed that periodic Floquet driving and engineered dissipation offer robust and experimentally accessible pathways to stabilize persistent collective oscillations beyond equilibrium paradigms \cite{SachaReview,annurevDTC,ColloquiumTC}.

In driven systems, discrete time crystals (DTCs) arise when observables exhibit subharmonic oscillations, breaking discrete time-translation symmetry  spontaneously~\cite{FloquetTC,FloquetLMG,DTCFloqetSondhi,DiscreteTCVishwanath,DTCLiu,GongDTC,Zhu2019,TimeCrystalOQS,DiscreteTCExperiment,DTCHuang,DTCHuang2,DTCsCaiZi,DTCsAchilleas,DTCDeng,DTCDeng2}. 
In contrast, CTCs appear in open quantum systems where coherent Hamiltonian competes with dissipation, giving rise to limit cycles in the long-time steady state \cite{BTC,BTCExperiment1,BTCExperiment2,BTCExperiment3,BTCSynchronization,QuantmTrajectoryBTC,CTCExperiment,CTCSolidState,CTCAtomCavity,CTCLongRangeLindbladian,CTCBeyondMeanField,CTCBeyondMeanField2,CTCCoupled,CTCCoupled2,BTCBeyondMeanField,CTCSingleModeCavity,CTCPeriodicDriven,VanDerPol}.  
These developments have revealed that time-crystalline order can naturally emerge in driven-dissipative settings governed by a Lindblad master equation. 

An even richer class of temporal organization occurs when the long-time response is quasi-periodic rather than strictly periodic, leading to the discovery of discrete quasi–time crystals (DQTCs) that are temporal analogs of spatial quasi-crystals \cite{QuasiperiodicalDrive1,QuasiperiodicalDrive2,QuasiperiodicalDrive3,QuasiperiodicalDrive4,DQTCZhaoHZ,DTCDTQCWu,DTQCRydberg,DTQCExperiment,CTCsQCTCs}. 
Such phases have been realized experimentally using drives with two incommensurate frequencies, for instance, in Rydberg-atom chains and nitrogen-vacancy (NV) spin ensembles \cite{DTQCExperiment2NVCenters,DTQCExperiment3}.  
However, these protocols typically require elaborate multi-frequency control or precise synchronization of incommensurate tones. A natural question thus arises: can quasi-periodic temporal order emerge under a single simple periodic drive, which is both of theoretical interest and experimental relevance?

In this Letter, we show that the answer is affirmative.  
We introduce a dissipative collective spin model~\cite{KickedTop1987,LiangCG,LMGDissipation,BTC,Passarelli2025chaosmagicin} , in which each spin couples identically to a common mode or collective decay channel.  
Our analysis demonstrates that this minimal setup—subjected only to a single periodic kick—can host a robust DQTCs phase, whose incommensurate frequencies are continuously tunable by the drive amplitude. This single-drive mechanism not only simplifies the generation of quasi-periodic temporal order, but also establishes a unified framework encompassing stationary, DTCs, DQTCs, and chaotic phases, providing a powerful and unified platform for realizing complex 
temporal orders in experimentally accessible systems.

\textit{Setup.—}
We consider a driven-dissipative collective model, described by the time-dependent Hamiltonian
\begin{equation}
\begin{aligned}
\hat{H}(t)&=\hat{H_0}+\hat{H}_1\sum_{n\in\Bbb{Z}}\delta(t-n)
\end{aligned}
\end{equation}
where $\hat{H}_{0}=\omega_{0}\hat{S}^{x}+(\omega_{z}/S)(\hat{S}^{z})^{2}$ represents the static part of the Hamiltonian, and $\hat{H}_{1}=\omega_{1}\hat{S}^{x}$ corresponds to an instantaneous kick applied at integer times.  
The collective spin operators $\hat{S}^{\alpha}=\tfrac{1}{2}\sum_{j=1}^{N}\hat{\sigma}_{j}^{\alpha}$ ($\alpha=x,y,z$).
The parameters $\omega_{0}$ and $\omega_{1}$ denote the strengths of the static transverse field and periodic kick, respectively, and $\omega_{z}$ controls the nonlinear collective interaction.  
The system thus describes an ensemble of $N$ identical two-level atoms or spins coupled collectively, characterized by the total spin $S=N/2$.  
Such all-to-all spin interactions are routinely achieved in experimental platforms, including atomic ensembles in optical cavities, superconducting qubits coupled to resonators, and large ensembles of NV centers or rare-earth ions interacting via collective photon modes \cite{RMPAtomsPhotons,RMPColdAtomsOptical,BTCExperiment1,BTCExperiment2,BTCExperiment3}

Dissipation in this system is introduced through a collective decay channel with a Lindblad operator
$\hat{L}=\sqrt{\kappa/S}\,\hat{S}^{-}$, yielding the Lindblad master equation~\cite{Lindblad,GKSL}
\begin{equation}
\label{eq:LME}
\frac{d\hat{\rho}}{dt}
=-\mathrm{i}[\hat{H}(t),\hat{\rho}]
+\frac{\kappa}{S}\!\left(\hat{S}^{-}\hat{\rho}\hat{S}^{+}
-\tfrac{1}{2}\{\hat{S}^{+}\hat{S}^{-},\hat{\rho}\}\right),
\end{equation}
where $\kappa/S$ is the collective dissipation rate and the collective raising and lowering are defined as $\hat{S}^{\pm}=\hat{S}^{x}\pm \mathrm{i}\hat{S}^{y}$.
The $1/S$ normalization ensures a well-defined thermodynamic limit.  
Throughout this work, we set $\kappa=1$.    
Since $\hat{S}^{2}$ commutes with both the Hamiltonian and the dissipation, the dynamics remain within the $S=N/2$ symmetric subspace of dimension $N+1$.

In the absence of the kick ($\omega_{1}=0$) and nonlinear term ($\omega_{z}=0$), Eq.~\eqref{eq:LME} reduces to a well-known boundary time crystal (BTC) model, exhibiting a transition from a stationary to a CTCs phase as $\omega_{0}/\kappa$ exceeds unity \cite{LarsonPRA,BTC}.  
Below the critical point, the system relaxes to a unique steady state, and above it, the dynamics approach a continuous family of non-isolated closed orbits on the Bloch sphere, corresponding to self-sustained oscillations of the collective magnetization.

\begin{figure}[ht]
\centering
\includegraphics[width=0.49\textwidth, keepaspectratio]{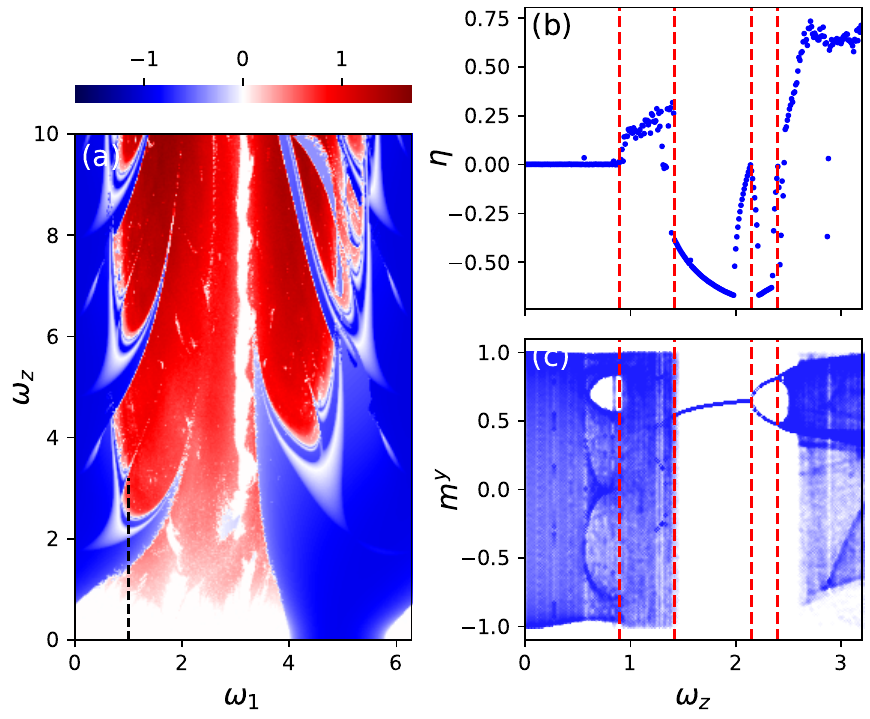}
\caption{(a) Largest Lyapunov exponent (LLE) $\eta$ in the semiclassical dynamics. $\omega_{0}=1.5$ and $\kappa=1.0$.
To compute the LLE, we evolve two initially nearby trajectories, $\vec{m}_1(0)=(0,0,1)$ and $\vec{m}_2(0)=\vec{m}_1(0)+\delta\vec{m}$, where the perturbation $\delta\vec{m}$ is random and very small ($|\delta\vec{m}|=10^{-8}$).
(b) LLE as a function of $\omega_z$ for fixed $\omega_1=1.0$, corresponding to the black dashed line in (a).
(c) The associated bifurcation diagram of $m^y$ along the same cut.
The red dashed lines at $\omega_z \approx 0.9,1.42,2.15,2.4$ indicate dynamical transitions.
As $\omega_z$ increases from $0$ to $3.2$, the system exhibits a sequence of dynamical regimes: limit-cycle (DQTC), chaos, a single fixed point, period-doubling with two stable points (DTC), and chaos again.}
\label{fig:LyapunovExponent}
\end{figure}

\begin{figure*}[htb]
\centering
\includegraphics[width=1.0\textwidth, keepaspectratio]{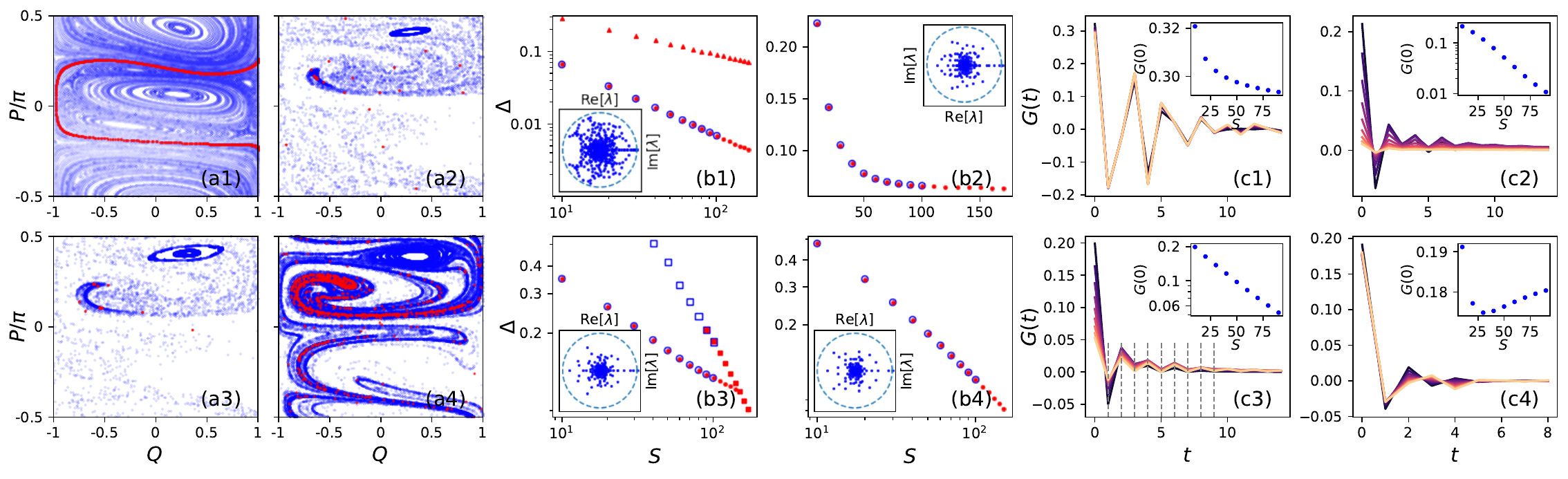}
\caption{The parameters are $\omega_{0}=1.5$, $\omega_{1}=1.0$, $\kappa=1.0$. Panels (a1)--(c1), (a2)--(c2), (a3)--(c3), and (a4)--(c4) correspond to $\omega_z=0.5$, $2.0$, $2.3$, and $3.0$, respectively.
(a1)--(a4) The Poincar\'e sections resulting from the mean-field equations. Red dots show the stroboscopic evolution of the initial state $\vec{m}(0)=(0,0,1)$, while blue dots show the stroboscopic evolution of 300 randomly sampled initial states. 
(b1)--(b4) Liouvillian gap and Liouvillian spectrum. 
Empty blue symbols correspond to exact diagonalization results, while filled red symbols indicate results obtained using the matrix-free Arnoldi method \cite{SUppMaterial}.
(b1) The upper triangles represent the gap of the Liouvillian eigenvalues with the smallest real part. 
(b3) Circles denote the smallest gap among Liouvillian eigenvalues with $\arg(\lambda)=0$, and rectangles denote the smallest gap with $\arg(\lambda)=\pi$. 
Insets plot the Floquet-Liouvillian spectrum.
(c1)--(c4) Connected autocorrelation function $G(t)$ for total spin ranging from $S=10$ (darkest) to $S=90$ (lightest).
Insets display the scaling of $G(0)$ with system size $S$. In (c3), dashed vertical lines mark integer times $t=1$--$9$.
}
\label{fig:PhasePortraitsLiouvillianGapAutoCorr}
\end{figure*}

\textit{Semi-classical picture.—}
To gain physical intuition, we analyze the semiclassical dynamics of the collective-spin model in the large-$S$ limit first, where quantum fluctuations are suppressed by $1/S$ (i.e., $\langle\hat{S}^{\alpha}\hat{S}^{\beta}\rangle\approx\langle\hat{S}^{\alpha}\rangle\langle\hat{S}^{\beta}\rangle$). 
Defining the normalized magnetization $\vec{m}=(m^{x},m^{y},m^{z})=\langle\hat{\bm{S}}\rangle/S$, the dissipative evolution between kicks obeys
\begin{equation}
\begin{aligned}
\dot m^{x} &= \kappa m^{x}m^{z}-2\omega_{z}m^{y}m^{z},\\
\dot m^{y} &= -\omega_{0}m^{z}+\kappa m^{y}m^{z}+2\omega_{z}m^{x}m^{z},\\
\dot m^{z} &= \omega_{0}m^{y}-\kappa[(m^{x})^{2}+(m^{y})^{2}],
\end{aligned}
\label{eq:meanfield}
\end{equation} 
At integer times, the instantaneous kick acts as a rigid rotation about the $x$ axis:
\begin{equation}
\begin{aligned}
m^{x\prime}&=m^{x},\quad
m^{y\prime}=m^{y}\cos\omega_{1}-m^{z}\sin\omega_{1},\\
m^{z\prime}&=m^{y}\sin\omega_{1}+m^{z}\cos\omega_{1},
\end{aligned}
\label{eq:kick}
\end{equation}
indicating that the dynamics are periodic in $\omega_{1}$ with period $2\pi$. The total spin length $|\vec{m}|$ is conserved, confining the motion to the Bloch sphere.  
We parameterize it via $(Q,P)$ as
$m^x=\sqrt{1-Q^2}\cos 2P$, $m^y=\sqrt{1-Q^2}\sin 2P$, and $m^z=Q$, which facilitates Poincar\'e analysis. Without loss of generality, we fix the initial state to be the fully polarized spin-up state throughout this work.

The dynamical character of the semiclassical motion is quantified by the largest Lyapunov exponent $\eta$, which measures the exponential divergence of nearby trajectories.  
For two initially close trajectories separated by $d_{0}$, their distance at time $t$ evolves as $d(t)\sim d_{0}e^{\eta t}$, yielding
\begin{equation}
\eta=\lim_{d_{0}\to0}\lim_{t\to\infty}\frac{1}{t}\ln\!\frac{d(t)}{d_{0}}.
\end{equation} 
Fig.~\ref{fig:LyapunovExponent}(a) shows the Lyapunov spectrum in the $(\omega_{z},\omega_{1})$ plane for $\omega_{0}=1.5$ and $\kappa=1.0$ with the initial state $\vec{m}=(0,0,1)$.  
White regions ($\eta=0$) correspond to the existence of closed orbits and signify the quasi-periodic oscillations. 
Blue regions ($\eta<0$) indicate stable fixed points and red regions ($\eta>0$) denote chaotic dynamics. 
Multiple arc-shaped bands with $\eta<0$ but very close to zero—visible as nearly white regions in the phase diagram—mark gradual crossover boundaries between different dynamical regimes.
The phase diagram features two main DQTC regions. 
At small $\omega_z$, DQTC behavior persists over a wide interval of $\omega_1$, with closed orbits arising for generic initial states.
A second regime emerges near $\omega_1\simeq\pi$, where quasi-periodic and chaotic trajectories coexist. Here, only a restricted set of initial conditions produces quasi-periodic motion, while most trajectories become chaotic, analogous to scar-like behavior~\cite{scarstadium,ScarExp,TurnerScar,TurnerScar2}.

Representative Poincar\'e sections are displayed in Figs.~\ref{fig:PhasePortraitsLiouvillianGapAutoCorr}(a1)-(a4), where the stroboscopic evolution of an initial state $\vec{m}(0)=(0,0,1)$ is also shown.
With fixed kick strength $\omega_{1}=1.0$, at $\omega_{z}=0.5$, the motion is regular on closed orbits.
At $\omega_z=2.0$, one stable fixed point emerges. Furthermore, there are two stable points for $\omega_z=2.3$, switching between them leading to period-doubling oscillations. At $\omega_z=3.0$, strange attractors appear, serving as a signature of dissipative chaos.

\textit{Floquet Lindbladian.—} The distinctive dynamical phases are also manifested in the Floquet Liouvillian spectrum and autocorrelation. 
The one-cycle time evolution map is $\mathcal{U}(\hat{\rho})=e^{-\mathrm{i}\hat{H}_1}e^{\mathcal{L}_0}(\hat{\rho})e^{\mathrm{i}\hat{H}_1}$, so that $\hat{\rho}(n)=\mathcal{U}\hat{\rho}(n-1)=\mathcal{U}^n\hat{\rho}(0)$. Let $\mathcal{U}\hat{\rho}_{j}=\lambda_{j}\hat{\rho}_{j}$ with eigenvalues ordered by magnitude  
$1=|\lambda_{0}|\ge|\lambda_{1}|\ge\cdots$.  
Here $\hat{\rho}_{0}\equiv\hat{\rho}_{\mathrm{ss}}$ denotes the steady state.  
Expanding the initial state as 
$\hat{\rho}(0)=\hat{\rho}_{\text{ss}}+\sum_{j}c_j\hat{\rho}_j$ gives
$\hat{\rho}(n)=\hat{\rho}_{\text{ss}}+\sum_{j\geq1}c_j(\lambda_j)^n\hat{\rho}_j$, and the distance to the steady state decays as $|\lambda_{1}|^{n}=e^{-n\Delta}$ after a sufficiently long time,  
which defines the Liouvillian gap $\Delta=-\log|\lambda_{1}|$.

In the DQTC regime [Fig.~\ref{fig:PhasePortraitsLiouvillianGapAutoCorr}(b1)], several complex eigenvalues approach the unit circle in the thermodynamic limit (TDL), with incommensurate phases. Their interference,  
$\hat{\rho}(n)\!\sim\!\sum_{j}c_{j}|\lambda_{j}|^{n}e^{\mathrm{i}n\arg\lambda_{j}}\hat{\rho}_{j}$,  
produces quasi-periodic oscillations.  
In the stationary phase [Fig.~\ref{fig:PhasePortraitsLiouvillianGapAutoCorr}(b2)], the Liouvillian spectrum is gapped, consistent with the finite Jacobian eigenvalues predicted semiclassically \cite{SUppMaterial}. 
In the period-doubling DTC phase [Fig.~\ref{fig:PhasePortraitsLiouvillianGapAutoCorr}(b3)], an eigenvalue with phase $\pi$ approaches the unit circle algebraically with $S$, yielding  
$\hat{\rho}(n)=\rho_{\mathrm{ss}}+(-1)^{n}c_{1}\hat{\rho}_{1}$ and thus period doubling.  
In the chaotic regime [Fig.~\ref{fig:PhasePortraitsLiouvillianGapAutoCorr}(b4)], the gap also closes algebraically; theory predicts a continuum of slowly decaying modes in the TDL \cite{CTCCoupled}, though this is hard to fully resolve at finite size.

The dynamical phases can also be diagnosed by the connected two-time correlation function  
\begin{equation}
    G(t) = \mathrm{Tr}\!\left[\,\tilde{\hat{S}}^{y}(t)\,\tilde{\hat{S}}^{y}(0)\,\hat{\rho}_{\mathrm{ss}}\,\right]/S^2,
\end{equation}
where  $\tilde{\hat{S}}^{y} = \hat{S}^{y} - \langle \hat{S}^{y} \rangle_{\mathrm{ss}}$ and $\langle\hat{S}^y\rangle_{\text{ss}}=\mathrm{Tr}[\hat{S}^y\hat{\rho}_{\text{ss}}]$. The operator evolves stroboscopically as $\hat{S}^y(t)=(\mathcal{U}^{\dagger})^t(\hat{S}^y)$, where $\mathcal{U}^{\dagger}$ denotes the Hermitian adjoint of $\mathcal{U}$ through $\mathrm{Tr}[\hat{A}\mathcal{U}(\hat{B})]=\mathrm{Tr}[\mathcal{U}^{\dagger}(\hat{A})\hat{B}]$ for arbitrary operators $\hat{A}$ and $\hat{B}$. 
Moreover, the steady-state fluctuation of $\hat{S}^y$ satisfy
\begin{equation}
F^{2} = [\langle (\hat{S}^{y})^{2} \rangle_{\mathrm{ss}} - \langle \hat{S}^{y} \rangle_{\mathrm{ss}}^{2}]/S = G(0)S, 
\end{equation}
so equal-time correlation $G(0)$ captures spatial fluctuations while $G(t)$ measures temporal order.

As shown in Fig.~\ref{fig:PhasePortraitsLiouvillianGapAutoCorr}(c1)-(c4), the lifetime of oscillations in $G(t)$ grows with $S$ in both the DQTCs and period-doubling DTCs phases, indicating persistent quasi-periodic or periodic dynamics in the TDL.  
By contrast, $G(t)\!\to\!0$ at long times in the stationary and chaotic phases.  
Moreover, $G(0)$ remains finite in the DQTCs and chaotic regimes but decays exponentially in the stationary phase, consistent with diverging versus bounded fluctuations.  
(For finite size, $G(0)$ may appear to decrease even in a DTCs regime, though it should remain finite in the TDL.)

These behaviors allow us to classify the four phases in the TDL:
(i) DQTCs: $G(0)\neq0$, $G(\infty)\neq0$, and $G(t)$ quasi-periodic;  
(ii) Stationary: $G(0)=G(\infty)=0$; 
(iii) DTCs: $G(0)\neq0$, $G(\infty)\neq0$, and $G(t)$ periodic;  
(iv) Chaotic: $G(0)\neq0$ but $G(\infty)=0$, 
similar to the correlation behavior of time glass, though the Liouvillian gap scaling differs \cite{hagatimeglasses}.

\begin{figure}[ht]
\centering
\includegraphics[width=0.5\textwidth, keepaspectratio]{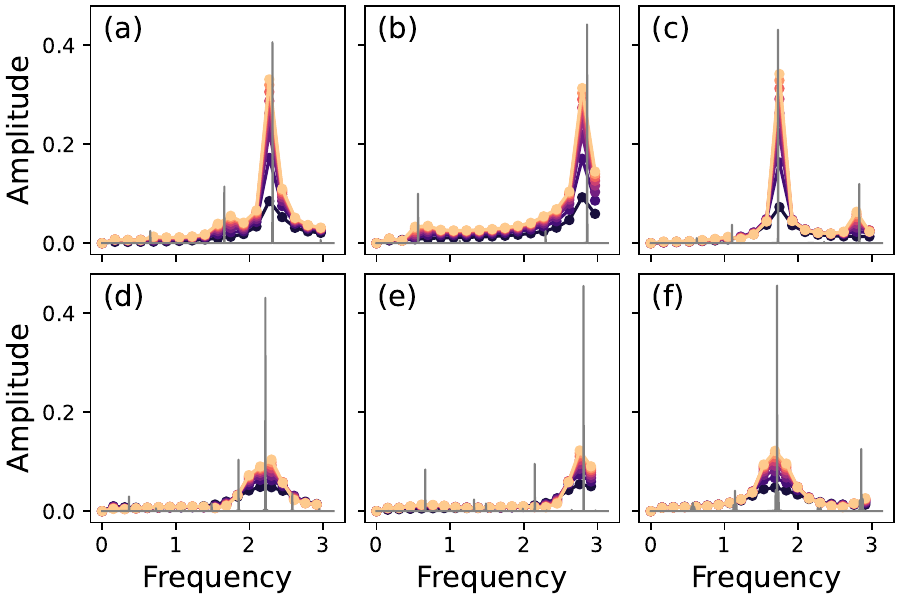}
\caption{Frequency response in the DQTC regime under varying kick strength. Curves correspond to system sizes from $S=10$ (darkest) to $S=80$ (lightest).
The gray line represents the Fourier spectrum of the corresponding semi-classical dynamics. $\omega_{0}=1.5, \kappa=1.0$. The upper row corresponds to $\omega_z=0$: (a) $\omega_1=1.0$, (b) $\omega_1=2.0$, and (c) $\omega_1=3.0$.
The lower row corresponds to $\omega_z=0.5$: (d) $\omega_1=1.0$, (e) $\omega_1=2.0$, and (f) $\omega_1=3.0$.  }
\label{fig:FrequencyShift}
\end{figure}

\textit{Frequency shift with kick strength.—}
As shown in Fig.~\ref{fig:LyapunovExponent}, for relatively small $\omega_z$, the closed periodic orbits survive for a wide range of kick strength $\omega_1$, giving rise to a robust DQTCs regime. We further show that the main frequency $\Omega_{\text{eff}}$ of the DQTCs can vary continuously with $\omega_1$.

We first analyze this frequency shift semiclassically. Assuming that the dynamics remain confined to a limit cycle in the presence of a weak kick. A perturbation $\delta\vec{m}$ (the kick) has two components: its projection along the tangent to the limit cycle produces a phase advance $\delta\theta$, whereas the normal component only displaces the trajectory transiently before it relaxes back without altering the phase to leading order. As shown in the Supplementary Material (SM) \cite{SUppMaterial}, this yields the stroboscopic frequency
\begin{equation}
\Omega_{\text{eff}}(\omega_1)\approx
\Omega+\omega_1\langle Z_{\text{geom}}\rangle_{\text{inv}}+\mathcal{O}(\omega_1^{2}),
\end{equation}
where the geometric phase-response factor $\langle Z_{\text{geom}}\rangle_{\text{inv}}$ is independent of $\omega_1$. Thus, $\Omega_{\text{eff}}$ shifts linearly with $\omega_1$ for weak kicks. Surprisingly, this linear dependence persists even for strong kicks, provided the limit cycle remains intact. Numerical simulations further show that this approximate linear shift survives over a range of $\omega_z$, even though the limit cycle becomes substantially deformed as $\omega_1$ increases \cite{SUppMaterial}.

Because the periodic kick is generally incommensurate with the intrinsic oscillation, the resulting dynamics are quasi-periodic and exhibit a Fourier spectrum containing peaks at
\begin{equation}\label{eq:FrequencyPeaks}
f = m\Omega_{\text{eff}} + n f_{\text{kick}}, \qquad m,n\in\mathbb{Z}.
\end{equation}
Fig.~\ref{fig:FrequencyShift} shows that the mean-field predictions agree well with finite-$S$ numerical simulations, and the SM demonstrates that Eq.~\eqref{eq:FrequencyPeaks} holds across a wide parameter range~\cite{SUppMaterial}.

A striking phenomenon emerges when the kick strength is tuned in some parameter window: in these regions $\Omega_{\text{eff}}$ becomes insensitive to both $\omega_1$ and $\omega_z$, and the ratio $\Omega_{\text{eff}}/f_{\text{kick}}$ locks to rational values such as $1/5$, $1/4$, $1/3$, $2/5$, or $1/2$ \cite{SUppMaterial}. Such a frequency locking pattern is often referred to as Arnold tongues in nonlinear dynamics \cite{Synchronizaton}, and we indeed identify the tongue-like structures in parameter space~\cite{SUppMaterial}. The resulting dynamics is similar to fractional DTC behaviors as the main frequency is now rational, though the Fourier spectrum also contains sidebands of the form $f=m\Omega_{\text{eff}}+nf_{\text{mod}}$, with an additional small modulation frequency $f_{\text{mod}}$.

In summary, within the DQTC regime, the effective stroboscopic frequency $\Omega_{\text{eff}}$ shifts approximately linearly with kick strength, except for fractional plateaus associated with Arnold tongues. The results demonstrate that the kick strength enables continuous tuning of quasi-periodic time-crystalline oscillations. Although the motion is quasi-periodic both inside and outside the locked regions, the distinct forms of the frequency spectra indicate different underlying mechanisms deserving further research.

\begin{figure}[ht]
\centering
\includegraphics[width=0.5\textwidth, keepaspectratio]{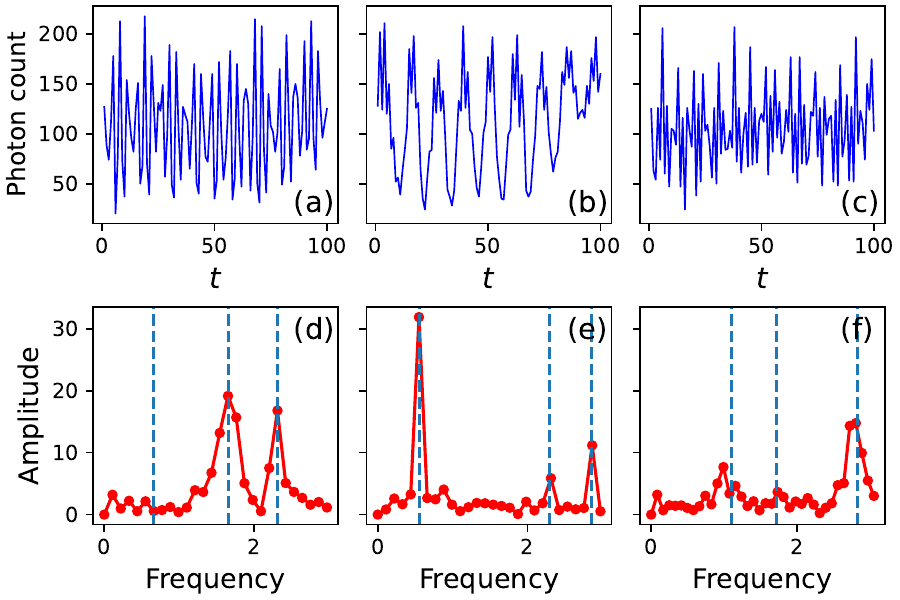}
\caption{Single quantum trajectory generated using the quantum jump method in the DQTCs regime. Parameters are fixed to $\omega_{0}=1.5$, $\kappa=1.0$, $\omega_{z}=0$ and $S=200$.
The three columns correspond to increasing kick strength: (left) $\omega_{1}=1.0$, (middle) $\omega_{1}=2.0$, and (right) $\omega_{1}=3.0$.
Top row: photon-count signal along the trajectory.
Bottom row: corresponding Fourier spectra of the photon-count time series.
Dashed blue lines indicate frequencies predicted from the semiclassical analysis. }
\label{fig:QuantumJump}
\end{figure}

\textit{Quantum trajectory and experimental realization.—}  
Thus far, we have analyzed the dynamics in the limits of large $S$ (mean-field approximation) and small $S$ (exact diagonalization). To access larger system sizes, we further unravel the Floquet Lindbladian using the quantum jump method, which expresses the density-matrix evolution as an ensemble of stochastic pure-state trajectories. Notably, quantum jumps have been directly observed in previous experiments \cite{QuantumJumpExperiment1,QuantumJumpExperiment2}. 

Figs.~\ref{fig:QuantumJump}(a)-(c) show the photon count for $\omega_1=1,\ 2,\ 3$ in a single quantum trajectory realization.  
The photon count records the timestamps of spontaneous emission photons, i.e., the instances when quantum jumps occur~\cite{BTCExperiment2}.
Surprisingly, the Fourier spectrum of a single trajectory already matches the semiclassical predictions [Figs.~\ref{fig:QuantumJump}(d)–(f)], demonstrating that the kick-induced frequency shift is accessible through photon-emission statistics.
This suggests a natural route for experimental realization. The BTC model ($\omega_{z}=0$, $\omega_{1}=0$) has already been implemented using a pencil-shaped cloud of laser-cooled atoms in free space \cite{BTCExperiment2}. Observing the tunable DQTC only requires adding a periodic kick $\hat{S}^{x}$. Importantly, the Raman field used for state preparation in Ref.~\cite{BTCExperiment2} already generates a transverse Hamiltonian proportional to $\hat{S}^x$; applying it as short, high-amplitude pulses once per Floquet cycle may realize this required kick.   
Other platforms of collective spin models \cite{BTCExperiment1,BTCExperiment3} can likewise be extended to implement our single-drive protocol.

\textit{Conclusions and discussions.---} 
In this Letter, we have investigated a driven-dissipative collective spin model and demonstrated that a single periodic drive is sufficient to generate continuously tunable DQTCs. Using semiclassical analysis, exact diagonalization, and quantum-jump simulations, we showed that the main frequency $\Omega_{\text{eff}}$ shifts approximately linearly with the kick strength, except on fractional plateaus where $\Omega_{\text{eff}}/f_{\text{kick}}$ becomes a fixed rational number.
The interplay between coherent dynamics, dissipation, and periodic driving further gives rise to a rich dynamical phase diagram encompassing DQTCs, stationary states, DTCs, and chaotic regimes.

We also demonstrated that the tunable DQTCs can be detected experimentally through photon-emission statistics. A straightforward extension of the experimental setup in Ref.~\cite{BTCExperiment2}, supplemented by a periodic kick, provides a feasible platform for realizing and observing this phenomenon. An interesting direction for future work is to generalize our analysis to quantum dissipative continuous time crystals, where the mean-field limit-cycle picture breaks down \cite{CTCBeyondMeanField,CTCBeyondMeanField2}.

\textit{Acknowledgement.---}  X.F. and S.C. were supported by the National Key Research
and Development Program of China (Grant No. 2023YFA1406704), the NSFC under Grants No. 12474287 
and No. T2121001. S.-X. Z. acknowledges the support by Quantum Science and Technology-National Science and Technology Major Project (2024ZD0301700). S. L. was supported by the Gordon and Betty Moore Foundation
through Grant No. GBMF8685 toward the Princeton theory
program, the Gordon and Betty Moore Foundation’s EPiQS
Initiative (Grant No. GBMF11070), the Office of Naval
Research (ONR Grant No. N00014-20-1-2303), the Global
Collaborative Network Grant at Princeton University,
the Simons Investigator Grant No. 404513, the BSF
Israel US foundation No. 2018226, the NSF-MERSEC (Grant No. MERSEC DMR 2011750), the Simons
Collaboration on New Frontiers in Superconductivity,
and the Schmidt Foundation at the Princeton University.

% \textit{Data availability.}  Numerical data for this manuscript are publicly accessible in Ref. \cite{data-available}.

\bibliography{ref}

\end{document}

\renewcommand{\thefigure}{S\arabic{figure}}
\setcounter{figure}{0}
\renewcommand{\theequation}{S\arabic{equation}}
\setcounter{equation}{0}
\renewcommand{\thesection}{\Roman{section}}
\setcounter{section}{0}
\setcounter{secnumdepth}{4}
\begin{center}
\textbf{\large Supplemental Material for ``Tunable discrete quasi-time crystal from a single drive''}
\end{center}

\author{Xu Feng}
\affiliation{Beijing National Laboratory for Condensed Matter Physics, Institute
of Physics, Chinese Academy of Sciences, Beijing 100190, China}
\affiliation{School of Physical Sciences, University of Chinese Academy of Sciences,
Beijing 100049, China }

\author{Shuo Liu}
% \email{sl6097@princeton.edu}
\affiliation{Department of Physics, Princeton University, Princeton, New Jersey 08544, USA}
\affiliation{Institute for Advanced Study, Tsinghua University, Beijing 100084, China}

\author{Shu Chen}
\email{schen@iphy.ac.cn}
\affiliation{Beijing National Laboratory for Condensed Matter Physics,
Institute of Physics, Chinese Academy of Sciences, Beijing 100190, China}
\affiliation{School of Physical Sciences, University of Chinese Academy of Sciences,
Beijing 100049, China }

\author{Shi-Xin Zhang}
\email{shixinzhang@iphy.ac.cn}
\affiliation{Beijing National Laboratory for Condensed Matter Physics, Institute
of Physics, Chinese Academy of Sciences, Beijing 100190, China}
\maketitle

\tableofcontents

\section{Mean-field equations}
We analyze the semiclassical (mean-field) dynamics of the collective spin. We define the rescaled spin components
\(
m^\alpha=\mean{\hat{S}^\alpha}/S
\)
with total spin length \(S\) (for \(N\) spin-1/2 particles, \(S=N/2\)). In the thermodynamic limit \(S\to\infty\), factorization
\(
\mean{m^\alpha m^\beta}\approx \mean{m^\alpha}\mean{m^\beta}
\)
is justified as \([m^\alpha,m^\beta]\to 0\).
The continuous-time evolution governed by the static Lindbladian \(\mathcal{L}_0\) yields
\begin{equation}
\label{eq:FirstPart}
\begin{aligned}
\dot m^{x} &= \kappa m^{x} m^{z}-2\omega_{z}m^y m^z, \\
\dot m^{y} &= -\omega_{0} m^{z} + \kappa m^{y} m^{z}+2\omega_{z}m^x m^z, \\
\dot m^{z} &= \omega_{0}m^y-\kappa \!\left[ (m^{x})^{2} + (m^{y})^{2} \right].
\end{aligned}
\end{equation}
A delta-kick generated by \(\hat{H}_1=\omega_1 \hat S^x\) instantaneously rotates \((m^y,m^z)\) around the \(x\)-axis:
\begin{equation}
\label{eq:SecondPart}
\begin{aligned}
{m^{x}}^{\prime}&=m^x, \\
{m^{y}}^{\prime}&=m^y\cos\omega_1-m^z\sin\omega_1, \\
{m^{z}}^{\prime}&=m^y\sin\omega_1+m^z\cos\omega_1.
\end{aligned}
\end{equation}

\paragraph*{No-kick case.}
Setting \(\omega_1=0\) and first \(\omega_z=0\), the dynamics exhibits a nonequilibrium transition at \(\omega_0/\kappa=1\). For \(\omega_0/\kappa<1\), the stable fixed point is
\(
(m^x,m^y,m^z)=(0,\omega_{0}/\kappa,-\sqrt{1-\omega_{0}^2/\kappa^2}).
\)
For \(\omega_0/\kappa>1\), trajectories are periodic and confined to constant
\(
m^2=(m^x)^2+(m^y)^2+(m^z)^2
\)
and
\(
\mathcal{M}=m^x/(m^y-\omega_0/\kappa)
\)
manifolds. For the initial condition \((0,0,1)\) used in the main text, one finds~\cite{QuantmTrajectoryBTC}
\begin{equation}\label{eq:SpinEvo}
m^x(t)=0,\quad
m^y(t)=\frac{\omega_0}{\kappa}+\frac{\omega_{0}^2/\kappa^2-1}{\cos(\Omega t-\phi_0)-\omega_{0}/\kappa},\quad
m^{z}(t)=\frac{\sqrt{\omega_{0}^2/\kappa^2-1}\,\sin(\Omega t-\phi_0)}{\cos(\Omega t-\phi_0)-\omega_{0}/\kappa},
\end{equation}
with \(\Omega=\sqrt{\omega_{0}^2-\kappa^2}\), \(\sin \phi_{0}=\sqrt{1-\kappa^2/\omega_{0}^2}\), and \(\cos\phi_{0}=\kappa/\omega_{0}\).
More specifically, $m^y(t)$ can be decomposed as the superposition of harmonic waves since 
\begin{equation}
\dfrac{1}{\cos(\Omega t-\phi_0)-\omega_0/\kappa}=-\cot(\phi_0)\left[1+2\sum_{n=1}^\infty r^n\cos(n(\Omega t-\phi_0))\right],
\end{equation}
in which $r=\dfrac{\omega_0}{\kappa}-\sqrt{(\dfrac{\omega_0}{\kappa})^2-1}$.
So the Fourier spectrum consists only of discrete peaks at 
$n\Omega$, and its amplitude decays as $r^n$. 

When \(\omega_z\neq 0\), \(\mathcal{M}\) is not conserved. An alternative invariant can be constructed~\cite{BTC}:
\begin{equation}
\begin{aligned}
\mathcal{R}_{\omega_z} &=
2 \omega_z \log \!\left[
\big( \kappa m^y + 2 \omega_z m^x - \omega_0 \big)^2
+ \big( \kappa m^x - 2 \omega_z m^y \big)^2
\right] \\
&\quad + 2\kappa \, \arctan \!\left[
\frac{\kappa m^x - 2 \omega_z m^y}
{\kappa m^y + 2 \omega_z m^x - \omega_0}
\right]
+ 2 \kappa \pi n,
\end{aligned}
\end{equation}
which reduces to \(\mathcal{R}_{0}=2\kappa\arctan \mathcal{M}\) for \(\omega_z=0\). These conserved quantities suggest that this system might exhibit closed periodic orbits in certain regimes, indicating the emergence of continuous time crystals.

\begin{figure*}[htb]
\centering
\includegraphics[width=1.0\textwidth]{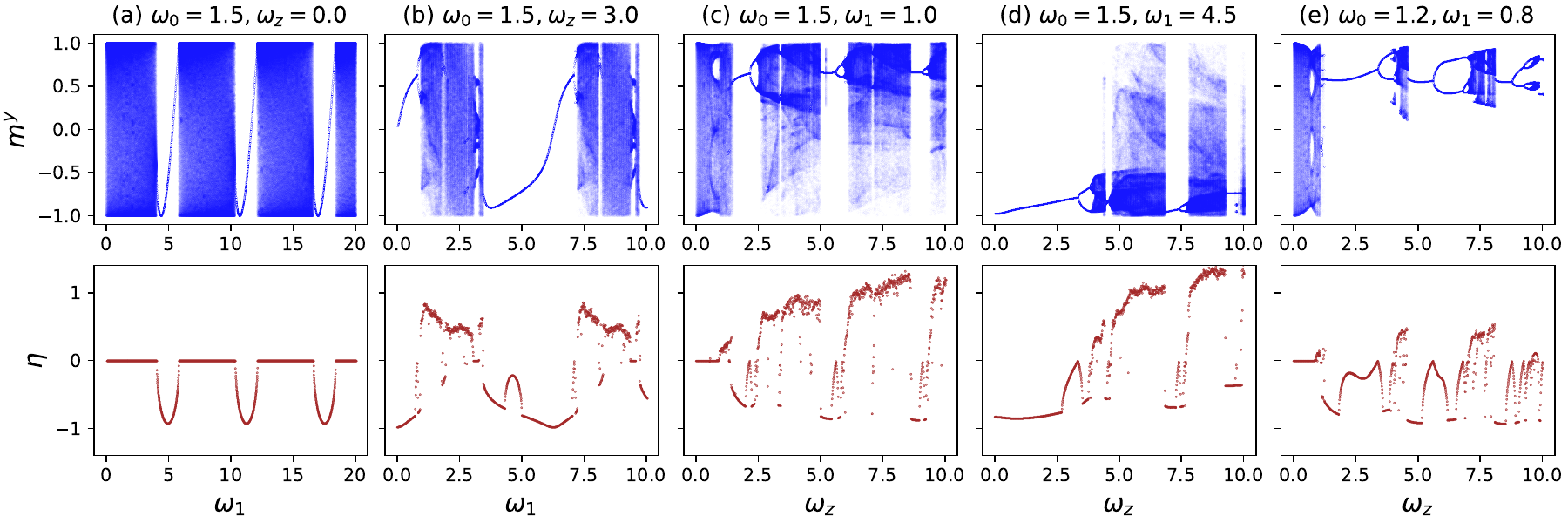}
\caption{(Top) Bifurcation diagrams of the mean-field dynamics. (Bottom) Largest Lyapunov exponents of the corresponding trajectories. We consider the two initial conditions $\vec{m}_1(t=0)=(0,0,1)$ and  $\vec{m}_2(t=0)=\vec{m}_1(t=0)+\delta \vec{m}$, in which the perturbation $\delta\vec{m}$ is random and very small ($\lVert\delta\vec{m}\rVert=10^{-8}$). }
\label{fig:bif_lyap}
\end{figure*}

\begin{figure*}[ht]
\centering
\includegraphics[width=1.0\textwidth, keepaspectratio]{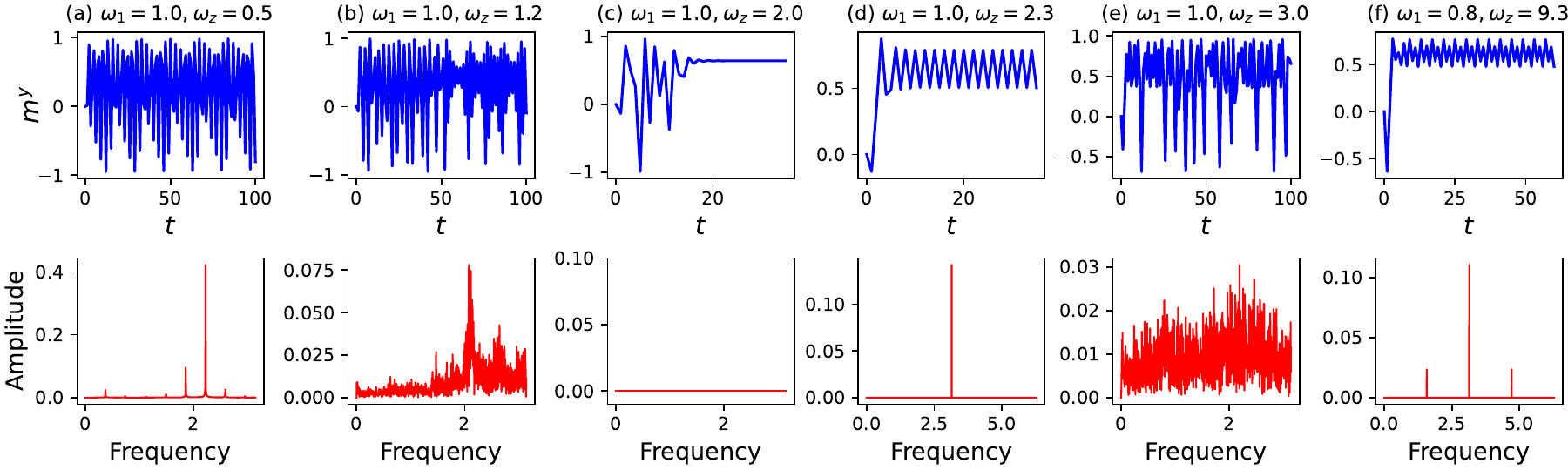}
\caption{Evolution of the magnetization $m^y$ (upper panels) and the corresponding Fourier spectra (lower panels) in the semiclassical limit. $\omega_{0}=1.5$, $\kappa=1.0$. (a) quasi-periodic oscillation; (b) chaos; (c) stable fixed point; (d) period-2 oscillation; (e) chaos; (f) period-4 oscillation. The initial state is $\vec{m}(0)=(0,0,1)$.}
\label{fig:SemiMyFourier}
\end{figure*}

\paragraph*{With kick.}
As shown in the main text, the largest Lyapunov exponent spectrum and Poincar\'e sections reveal four distinct dynamical regimes.  
In addition, Fig.~\ref{fig:bif_lyap} displays the bifurcation diagrams together with the Lyapunov exponents as functions of the kick strength $\omega_1$ and nonlinear interaction $\omega_z$, illustrating how the dynamics change as these parameters are varied.  
Fig.~\ref{fig:SemiMyFourier} further shows $m^y(t)$ and its Fourier spectrum for representative parameter choices, highlighting the qualitative differences between quasi-periodic, periodic, stationary, and chaotic dynamics.

\section{Jacobian matrix}
Because the fixed point of the kicked collective spin model cannot be obtained analytically, and the nonlinear structure of Eq.~\eqref{eq:FirstPart} prevents a direct linearized analytical approximation around the fixed points, we evaluate the Jacobian matrix numerically. We construct the Jacobian matrix from the one-period stroboscopic map. One driving period consists of (i) dissipative mean-field evolution governed by Eq.~\eqref{eq:FirstPart} for a duration $T=1$, followed by (ii) an instantaneous rotation of the Bloch vector around the $x$ axis described by Eq.~\eqref{eq:SecondPart}. Denoting the continuous evolution by $\mathcal{U}_T(\vec{m})$ and the kick by $\mathcal{U}_K(\vec{m})$, the Floquet map is given by
\begin{equation}
    \vec{m}_{n+1} = \mathcal{U}(\vec{m}_n) = \mathcal{U}_K\!\big[\mathcal{U}_{T}(\vec{m}_n)\big].
\end{equation}

The fixed point $\vec{m}_*$ is of the kicked dynamics is then obtained numerically by iterating $\mathcal{U}$  until convergence. To evaluate the Jacobian of the map, we compute
\begin{equation}
    J = \left.\frac{\partial \mathcal{U}(\vec{m})}{\partial \vec{m}}\right|_{\vec{m}=\vec{m}_*},
\end{equation}
using a central finite-difference scheme, where each column corresponds to a perturbed initial condition propagated through one full Floquet cycle. Diagonalization of $J$ yields three eigenvalues $\{\mu_j\}$, whose magnitudes determine whether perturbations decay or grow.  The eigenvalues $\{\mu_j\}$ of $J$ characterize the local dynamics: the fixed point is stable when all eigenvalues satisfy $|\mu_j|<1$, and becomes unstable once any eigenvalue crosses the unit circle and satisfies $|\mu_j|>1$.
For fixed points, the relevant eigenvalue of the Jacobian matrix equals the Liouvillian gap in the thermodynamic limit.
Furthermore, the corresponding local Lyapunov exponents can also be obtained as
\begin{equation}
    \eta_j = \frac{1}{T}\ln |\mu_j|,
\end{equation}

For example, for the parameters $\omega_0=1.5$, $\kappa=1.0$, $\omega_1=0$, and $\omega_z=2.0$, the iterated map gives the fixed point $\vec{m}_*\approx(0.437,0.641,-0.631)$. The Jacobian matrix has been numerically constructed, whose relevant eigenvalue is $-0.649$ and its corresponding Lyapunov exponent is $-0.432$.
The largest exponent agrees with the Lyapunov exponent extracted from long-time semiclassical trajectories shown in Fig.~1 of the main text, confirming the validity of the Jacobian analysis. Moreover, the dominant contracting eigenvalue $-0.649$ is consistent with the Liouvillian spectral gap obtained from exact diagonalization of the Floquet Lindbladian [see Fig. (b2) in the main text], confirming that the Liouvillian gap remains finite for the stationary phase in the thermodynamic limit.

\section{Phase portraits of quasi-time crystal phase}
\begin{figure*}[htb]
\centering
\includegraphics[width=1.0\textwidth]{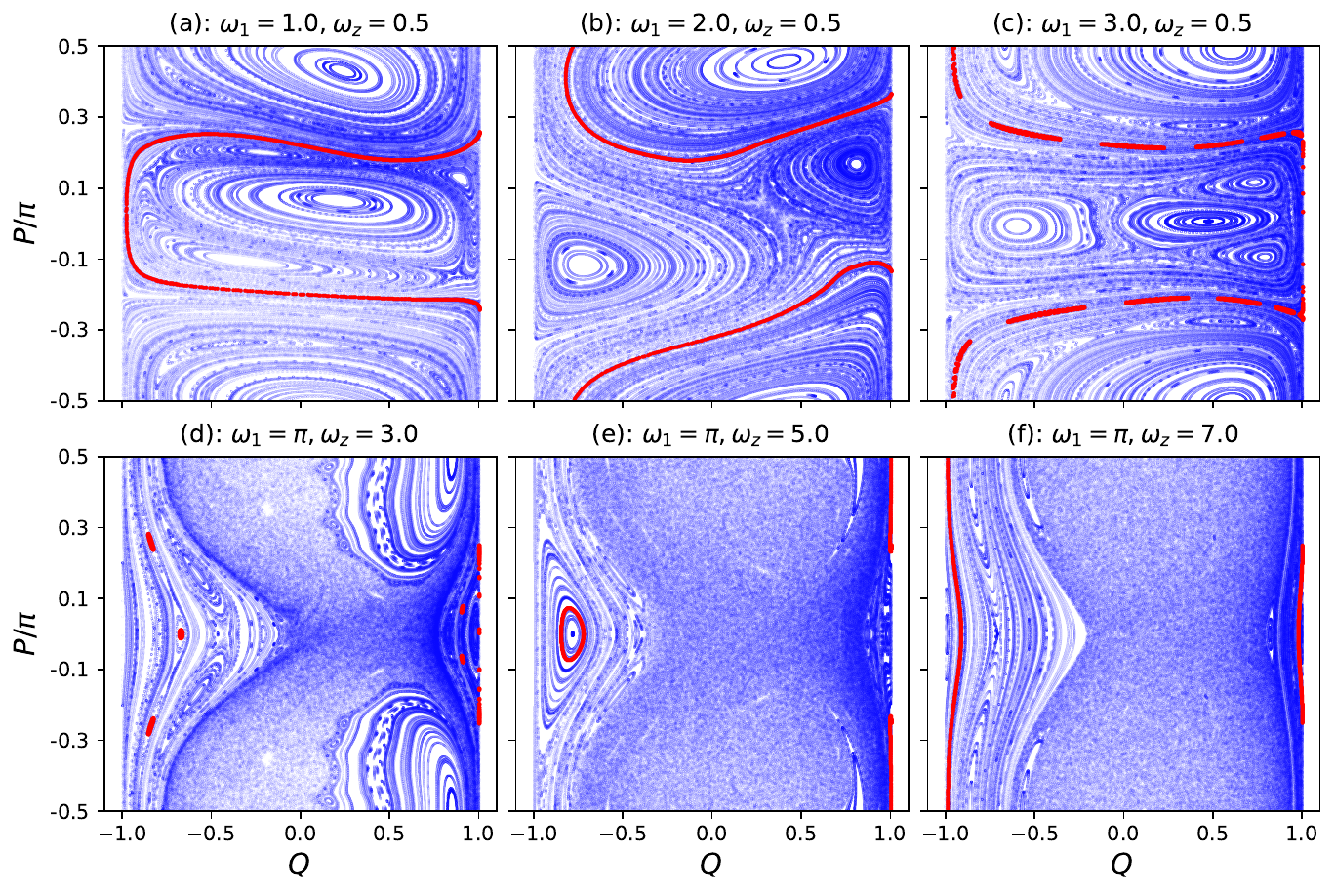}
\caption{Poincar\'e sections obtained from the mean-field equations.
$\omega_{0}=1.5$, $\kappa=1.0$. Red dots show the stroboscopic evolution of the initial state $\vec{m}(0)=(0,0,1)$.}
\label{fig:QuasiTimeOscillationPhasePortraits}
\end{figure*}

As discussed in the main text, the DQTCs phase is mainly distributed in two regions of parameter space.  
Region I is located at the lower-left part of the phase diagram, where $\omega_z$ is small, while Region II is centered around $\omega_1=\pi$.  
These two regions have distinct dynamical structures.  
As shown in Fig.~\ref{fig:QuasiTimeOscillationPhasePortraits}, the phase-space portrait in Region I  consists solely of closed periodic orbits, implying quasi-periodic oscillations for essentially all initial conditions.  
By contrast, in Region II, the phase space exhibits a coexistence of closed orbits and chaotic trajectories: quasi-periodic oscillations persist only for certain special initial states, whereas generic initial conditions display chaotic behavior.  
Interestingly, this structure resembles the phenomenology of quantum scars: although the global phase space is dominated by ergodic, thermal behavior, a restricted set of initial conditions supports long-lived, nonthermal trajectories \cite{scarstadium,ScarExp,TurnerScar,TurnerScar2}.

\section{Theoretical analysis of the semi-classical dynamics}
\begin{figure}[ht]
\centering
\includegraphics[width=1.0\textwidth, keepaspectratio]{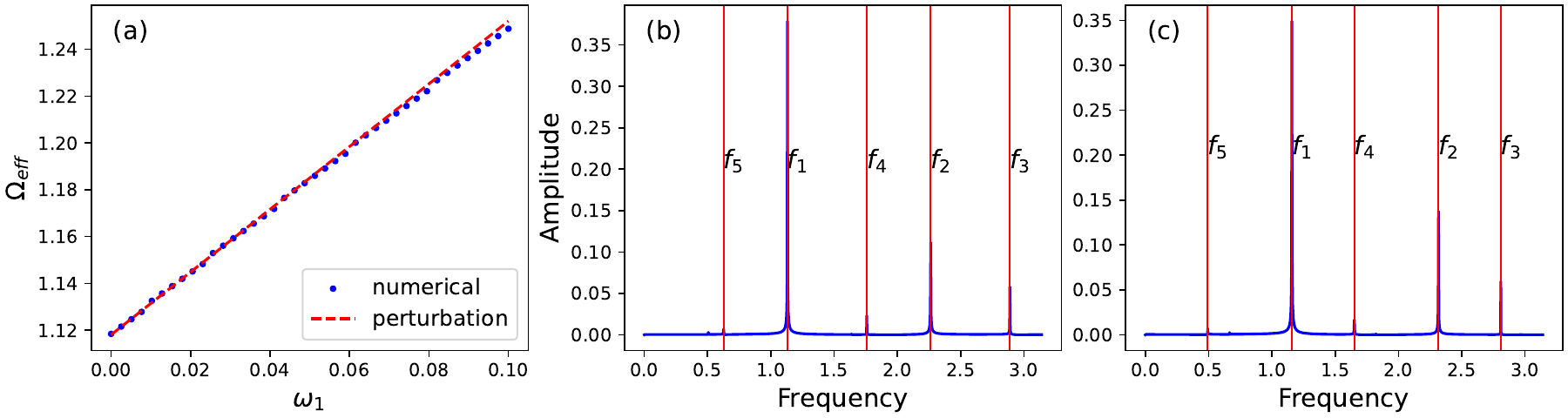}
\caption{
Frequency shift for very small $\omega_1$. The initial state is $\vec{m}(0)=(0,0,1)$.
(b) $\omega_1=0.01$: the peak frequencies are $f_1\approx1.131=\Omega_{\text{eff}}$, $ f_2\approx2.263=2\Omega_{\text{eff}}$, $f_3\approx2.889=-3\Omega_{\text{eff}}+f_{\text{kick}}$, $f_4\approx1.758=-4\Omega_{\text{eff}}+f_{\text{kick}}$, and $f_5\approx0.626=-5\Omega_{\text{eff}}+f_{\text{kick}}$.  
(c) $\omega_1\approx0.03$: the peak frequencies are $f_1\approx1.158=\Omega_{\text{eff}}$, $f_2\approx2.316=2\Omega_{\text{eff}}$, $f_3\approx2.809=-3\Omega_{\text{eff}}+f_{\text{kick}}$, $f_4\approx1.652=-4\Omega_{\text{eff}}+f_{\text{kick}}$, and $f_5\approx0.493=-5\Omega_{\text{eff}}+f_{\text{kick}}$. }
\label{fig:FrequencySpectrum}
\end{figure}
\subsection{Frequency shift for very small kick}
As illustrated above, in the absence of kicks, the steady-state semiclassical dynamics is characterized by a self-sustained limit cycle with internal frequency $\Omega=\sqrt{\omega_0^2-\kappa^2}$.  
The dynamics along this limit cycle are described by
\begin{equation}
\begin{aligned}
m^y(t)&=\dfrac{\omega_0}{\kappa}+\dfrac{\omega_0^2/\kappa^2-1}{\cos\theta-\omega_0/\kappa}, \\
m^z(t)&=\dfrac{\sqrt{\omega_0^2/\kappa^2-1}\sin\theta}{\cos\theta-\omega_0/\kappa},
\end{aligned}
\end{equation}
in which $\theta=\Omega t-\phi_0$. 

We now consider the effect of a weak periodic kick and assume that the dynamics remain confined to the limit cycle.  
The key idea is that a small perturbation $\delta \vec{m}$ decomposes into a component tangent to the limit cycle and a component normal to it.  
The tangential component ``pushes" the system forward along the trajectory, corresponding to a phase advance $\delta\theta$, whereas the normal component displaces the trajectory away from the limit cycle but rapidly relaxes back without changing the phase to leading order.

For small $\omega_1$, the kick modifies the magnetization to first order as
\begin{equation}
\begin{aligned}
    \delta m^y&=m^y_{\text{after}}-m^y_{\text{before}}=-\omega_1 m^z, \\
    \delta m^z&=m^z_{\text{after}}-m^z_{\text{before}}=\omega_1 m^y.
\end{aligned}
\end{equation}
Thus, the infinitesimal displacement due to one kick is
\begin{equation}
\delta\vec{m}=\omega_1(-m^z,m^y).
\end{equation}
Along the unperturbed limit cycle $\vec{m}(\theta)=(m^y(\theta),m^z(\theta))$, the tangent vector is 
\begin{equation}
    \vec{T}(\theta)=\dfrac{d\vec{m}}{dt}=(\dot{m}^y,\dot{m}^z).
\end{equation}
The kick shifts the state by $\delta \vec{m}$, and its projection along the tangent produces an instantaneous phase advance
\begin{equation}
\delta\theta(\theta)=\frac{\vec T(\theta)\!\cdot\!\delta\vec{m}}{\|\vec T(\theta)\|^2}\Omega=\omega_1\frac{\vec T(\theta)\!\cdot\!(-m^z(\theta),m^y(\theta))}{\|\vec T(\theta)\|^2}\Omega
\equiv \omega_1 Z_{\text{geom}}(\theta),
\end{equation}
where $Z_{\text{geom}}(\theta)$ is the geometric phase-response function.
After one period $T_{\text{kick}}=1$, the phase advances by
\begin{equation}
\theta_{n+1}=\theta_n+\Omega+\omega_1 Z_{\text{geom}}(\theta_n)+O(\omega_1^2),
\end{equation}
which is dubbed the circle map \cite{CircleMapI1984,CircleMapII1984}. So that the observed stroboscopic frequency is
\begin{equation}
\Omega_{\text{eff}}(\omega_1)=
\Omega+\omega_1\langle Z_{\text{geom}}\rangle_{\text{inv}}+O(\omega_1^2),
\end{equation}
where the invariant (residence-time-weighted) average is
\begin{equation}
\langle Z_{\text{geom}}\rangle_{\text{inv}}=
\frac{\displaystyle\int_0^{2\pi}\! Z_{\text{geom}}(\theta)/\dot\theta(\theta)\,d\theta} {\displaystyle\int_0^{2\pi}\! 1/\dot\theta(\theta)\,d\theta}.
\end{equation}
Since $\dot\theta(\theta)$ is nonuniform in the presence of the kick, the system spends more time in slow regions, enhancing the average response.
A numerical evaluation yields
\begin{equation}
\langle Z_{\text{geom}}\rangle_{\text{inv}}\approx1.34,
\end{equation}
so that $\Omega_{\text{eff}}\approx\Omega+1.34*\omega_1$ for small $\omega_1$.

For such small kicks, the limit cycle is not destroyed, and the dynamics remain quasi-periodic.  
As depicted in Fig.~\ref{fig:FrequencySpectrum}, the peaks of the frequency spectrum satisfy
$f=m\Omega_{\text{eff}}+nf_{\text{kick}}$
with integers $m,n\in\mathbb{Z}$.  
In our model: (i) the limit cycle oscillates with intrinsic frequency $\Omega_{\text{eff}}$; (ii) the kick is applied periodically once per unit time, corresponding to the external drive frequency $f_{\text{kick}}=2\pi$.  
The observable $m^y(t)$ after many kicks contains contributions from both oscillations.  
If these frequencies are incommensurate, the resulting signal is quasi-periodic, with $m\Omega_{\text{eff}}$ representing the internal harmonics of the self-sustained oscillation. The periodic kicks then act to shift and mix these internal frequencies.

\subsection{General case }
\begin{figure}[ht]
\centering
\includegraphics[width=1.0\textwidth, keepaspectratio]{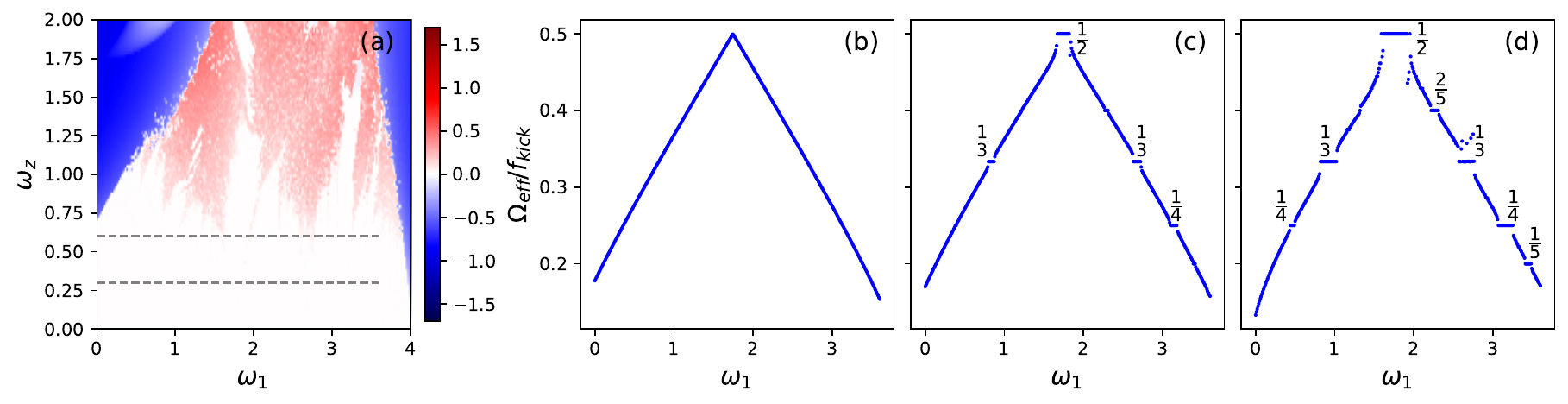}
\caption{Frequency shift with kick strength and Devil's
staircase. The initial state is $\vec{m}(0)=(0,0,1)$. The parameters are chosen as $\omega_0=1.5$, $\kappa=1.0$. (a) The largest Lyapunov exponent. The three dashed lines represent the parameters of (b)-(d). 
(b) $\omega_z=0.0$; (c) $\omega_z=0.3$; (d) $\omega_z=0.6$. 
The effective frequency $\Omega_{\text{eff}}$ varies approximately linearly with $\omega_1$ except for fractional plateaus.  
The Fourier spectrum is symmetric around $\pi$: for every peak at frequency $f$, there is a corresponding peak at $2\pi-f$.}
\label{fig:FrequencyShiftNumerical}
\end{figure}

For small $\omega_z$, the limit cycles persist even for relatively large $\omega_1$, implying that quasi-periodic oscillations survive over an extended parameter range.
In general, for example, the kick strength is large and the non-linear interaction $\omega_z\neq0$, it is found that the expected frequency response still holds $f=m\Omega_{\text{eff}}+nf_{\text{kick}}$ in a wide range of parameters. In the following, we present the numerical evidence supporting the frequency structure for various parameters. The frequencies $f_i$ below are indexed in descending order of their peak height in the Fourier spectrum. 

For $\omega_z=0$:
\begin{itemize}
\item[(1.1)] $\omega_1=1.0$:
\(
f_1\approx2.312=\Omega_{\text{eff}},\,
f_2\approx1.659=-2\Omega_{\text{eff}}+f_{\text{kick}},\,
f_3\approx0.653=3\Omega_{\text{eff}}-f_{\text{kick}},\,
f_4\approx2.966=4\Omega_{\text{eff}}-f_{\text{kick}}.
\)
\item[(1.2)] $\omega_1=2.0$:
\(
f_1\approx2.859=\Omega_{\text{eff}},\,
f_2\approx0.564=-2\Omega_{\text{eff}}+f_{\text{kick}},\,
f_3\approx2.295=3\Omega_{\text{eff}}-f_{\text{kick}},\,
f_4\approx1.130=-4\Omega_{\text{eff}}+2f_{\text{kick}}.
\)
\item[(1.3)] $\omega_1=3.0$:
\(
f_1\approx1.727=\Omega_{\text{eff}},\,
f_2\approx2.829=-2\Omega_{\text{eff}}+f_{\text{kick}},\,
f_3\approx1.102=-3\Omega_{\text{eff}}+f_{\text{kick}}\ ,
f_4\approx0.624=4\Omega_{\text{eff}}-f_{\text{kick}},\,
f_5\approx2.352=5\Omega_{\text{eff}}-f_{\text{kick}}.
\)
\end{itemize}

For nonzero $\omega_z$, such as $\omega_z=0.5$:

\begin{itemize}
\item[(2.1)] $\omega_1=1.0$:
\(
f_1\approx2.216=\Omega_{\text{eff}},\,
f_2\approx1.851=-2\Omega_{\text{eff}}+f_{\text{kick}},\,
f_3\approx2.581=4\Omega_{\text{eff}}-f_{\text{kick}},\,
f_4\approx0.365=3\Omega_{\text{eff}}-f_{\text{kick}},\,
f_5\approx1.486=-5\Omega_{\text{eff}}+2f_{\text{kick}},\,
f_6\approx0.730=6\Omega_{\text{eff}}-2f_{\text{kick}}.
\)
\item[(2.2)] $\omega_1=2.0$:
\(
f_1\approx2.810=\Omega_{\text{eff}},\,
f_2\approx2.148=3\Omega_{\text{eff}}-f_{\text{kick}},\,
f_3\approx0.662=-2\Omega_{\text{eff}}+f_{\text{kick}},\,
f_4\approx1.325=-4\Omega_{\text{eff}}+2f_{\text{kick}},\,
f_5\approx1.486=5\Omega_{\text{eff}}-2f_{\text{kick}},\,
f_6\approx1.987=-6\Omega_{\text{eff}}+3f_{\text{kick}}.
\)
\item[(2.3)] $\omega_1=3.0$:
\(
f_1\approx1.713=\Omega_{\text{eff}},\,
f_2\approx2.856=9\Omega_{\text{eff}}-2f_{\text{kick}},\,
f_3\approx1.143=-3\Omega_{\text{eff}}+f_{\text{kick}},\,
f_4\approx0.571=4\Omega_{\text{eff}}-f_{\text{kick}},\,
f_5\approx2.284=5\Omega_{\text{eff}}-f_{\text{kick}}.
\)
\end{itemize}

For the above cases, once $\Omega_{\text{eff}}$ is known, the possible positions of the frequency peaks can be inferred.  
The circle-map analysis in the small-$\omega_1$ limit predicts a linear increase of $\Omega_{\text{eff}}$ with $\omega_1$.  
Remarkably, Fig.~\ref{fig:FrequencyShiftNumerical} shows that this approximate linear dependence persists even for larger $\omega_1$.  
The slope is somewhat reduced compared to the small-$\omega_1$ prediction, likely due to higher-order corrections.  
Moreover, as seen in Fig.~\ref{fig:QuasiTimeOscillationPhasePortraits}, the shape of the limit cycle becomes strongly distorted at large $\omega_1$, making it an interesting open question why $\Omega_{\text{eff}}$ continues to shift nearly linearly with the kick strength.
However, as shown in Fig.~\ref{fig:FrequencyShiftNumerical}(c) and (d), there are plateaus for relatively large $\omega_z$, in which $\Omega_{\text{eff}}$ is invariant with $\omega_1$. In such locked regions, we will show that the response frequency will no longer obey $m\Omega_{\text{eff}}+nf_{\text{mod}}$.

\subsection{Locked $\Omega_{\text{eff}}$}
\begin{figure*}[ht]
\centering
\includegraphics[width=1.0\textwidth, keepaspectratio]{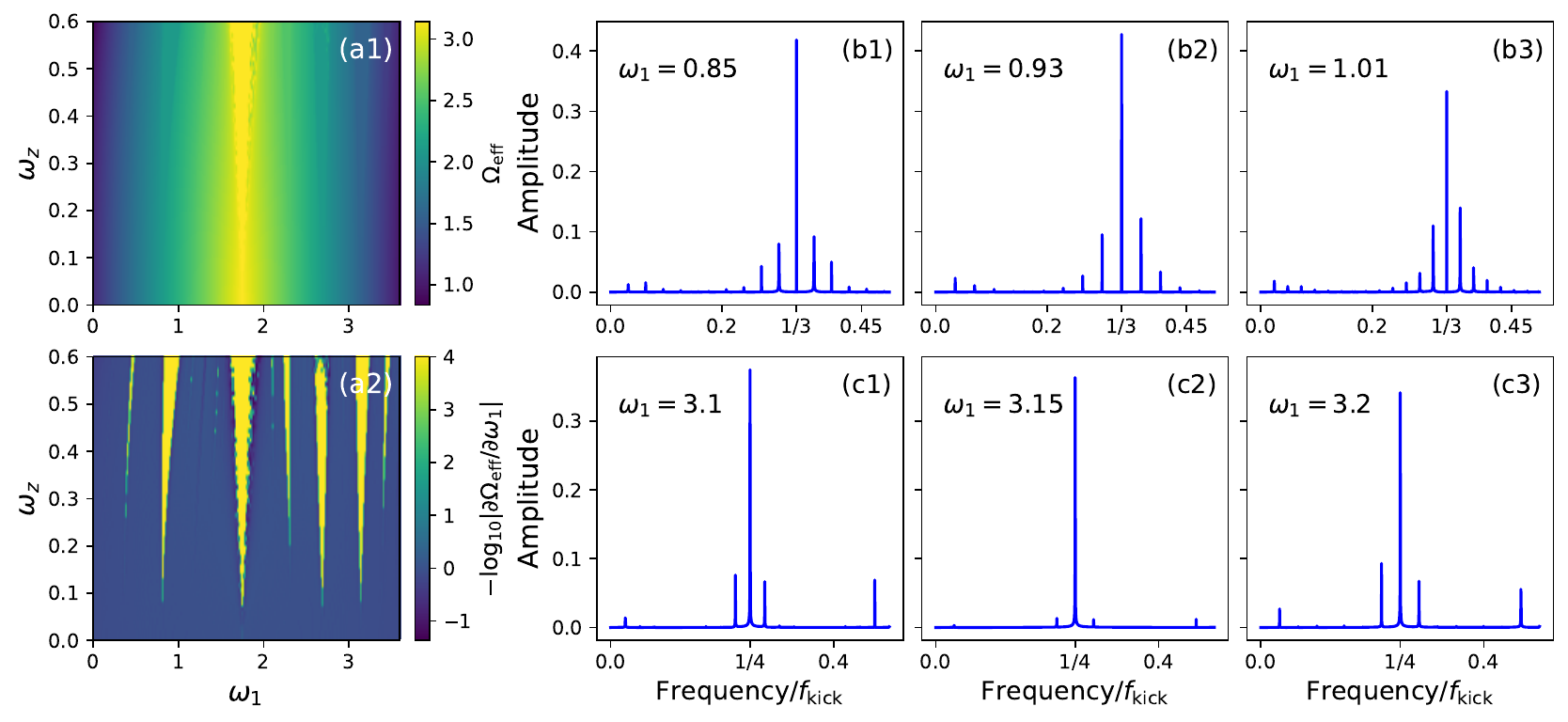}
\caption{Arnold tongue and locked frequency. The initial state is $\vec{m}(0)=(0,0,1)$, with parameters $\omega_0=1.5$, $\kappa=1.0$. For (b1)-(b3) and (c1)-(c3) $\omega_z=0.6$. (a1) Main stroboscopic frequency $\Omega_{\text{eff}}$. (a2) Arnold's  tongue diagram showing regions where $\Omega_{\text{eff}}/f_{\text{kick}}$ locks to rational values. (b1)-(b3) Fourier spectra in the $1/3$ locked region. (c1)-(c3) Fourier spectra in the $1/4$ locked region.}
\label{fig:LockedFrequency}
\end{figure*}
Surprisingly, we find parameter regimes in which the effective frequency $\Omega_{\text{eff}}$ remains unchanged when varying either $\omega_1$ or $\omega_z$. In these regions, the system locks to rational fractions of the drive, satisfying
$\Omega_{\text{eff}}/f_{\text{kick}}=p/q,\quad p, q\in\Bbb{Z}$.
As shown in Fig.~\ref{fig:LockedFrequency}(a2), these plateaus extend continuously in $\omega_z$, forming tongue-like structures characteristic of Arnold tongues in nonlinear dynamics \cite{Synchronizaton}.
Although one might expect strictly periodic motion whenever $\Omega_{\text{eff}}/f_{\text{kick}}$ is rational, the dynamics also feature additional modulation frequencies. In addition to harmonics of the locked frequency $m\Omega_{\text{eff}}$, the Fourier spectrum exhibits a small modulation frequency $f_{\text{mod}}$, and the observed peaks follow
\begin{equation}  f=m\Omega_{\text{eff}}+nf_{\text{mod}}, \quad m,n\in\Bbb{Z}.  
\end{equation}
Representative spectra illustrating this frequency organization are shown in Fig.~\ref{fig:LockedFrequency}(b1)–(b3) and (c1)–(c3).

For $\omega_z=0.6$ (The following equations are within error tolerance $O(10^{-3})$.)

$1/3$ locked case.

\begin{itemize}
\item[(1.1)] $\omega_z=0.85$: 
\(f_1=\Omega_{\text{eff}}\approx2\pi/3,\ f_2\approx2.2922=\Omega_{\text{eff}}+f_{\text{mod}},\ f_3\approx1.8966=\Omega_{\text{eff}}-f_{\text{mod}},\ f_4\approx2.4895=\Omega_{\text{eff}}+2f_{\text{mod}},\ f_5\approx1.6992=\Omega_{\text{eff}}-2f_{\text{mod}},\ f_6\approx0.3951=2f_{\text{mod}},\ f_7\approx0.1978=f_{\text{mod}},\ f_8\approx2.6876=\Omega_{\text{eff}}+3f_{\text{mod}},\ f_9\approx1.5012=\Omega_{\text{eff}}-3f_{\text{mod}},\ 
f_{10}\approx2.8847=\Omega_{\text{eff}}+4f_{\text{mod}}
.
\)
\item[(1.2)] $\omega_z=0.93$:
\(f_1=\Omega_{\text{eff}}\approx2\pi/3,\ f_2\approx2.3136=\Omega_{\text{eff}}+f_{\text{mod}},\ f_3\approx1.8752=\Omega_{\text{eff}}-f_{\text{mod}},\ f_4\approx2.5328=\Omega_{\text{eff}}+2f_{\text{mod}},\ f_5\approx1.6560=\Omega_{\text{eff}}-2f_{\text{mod}},\ f_6\approx0.2192=f_{\text{mod}}\ f_7\approx0.4384=2f_{\text{mod}},\ f_8\approx2.7520=\Omega_{\text{eff}}+3f_{\text{mod}},\ f_9\approx1.4368=\Omega_{\text{eff}}-3f_{\text{mod}},\ 
f_{10}\approx0.6576=3f_{\text{mod}}.
\)
\item[(1.3)] $\omega_z=1.01$:
\(f_1\approx2\pi/3=\Omega_{\text{eff}},\ f_2\approx2.2465=\Omega_{\text{eff}}+f_{\text{mod}},\ f_3\approx1.9423=\Omega_{\text{eff}}-f_{\text{mod}},\ f_4\approx2.3976=\Omega_{\text{eff}}+2f_{\text{mod}},\ f_5\approx1.7912=\Omega_{\text{eff}}-2f_{\text{mod}},\ f_6\approx2.5496=\Omega_{\text{eff}}+3f_{\text{mod}},\ f_7\approx0.1521=f_{\text{mod}},\ f_8\approx1.6392=\Omega_{\text{eff}}-3f_{\text{mod}},\ f_9\approx0.3032=2f_{\text{mod}},\ f_{10}\approx0.4552=3f_{\text{mod}}.
\)
\end{itemize}

$1/4$ locked case.

\begin{itemize}
\item[(2.1)] $\omega_z=3.1$: 
\(f_1=\Omega_{\text{eff}}\approx\pi/2,\ f_2\approx1.4060=\Omega_{\text{eff}}-f_{\text{mod}},\ f_3\approx2.9768=2\Omega_{\text{eff}}-f_{\text{mod}},\ f_4\approx1.7369=\Omega_{\text{eff}}+f_{\text{mod}},\ f_5\approx0.1653=f_{\text{mod}}.
\)
\item[(2.2)] $\omega_z=3.15$:
\(f_1=\Omega_{\text{eff}}\approx\pi/2,\ f_2\approx1.3641=\Omega_{\text{eff}}+f_{\text{mod}},\ f_3\approx2.9350=2\Omega_{\text{eff}}-f_{\text{mod}},\ f_4\approx1.7788=\Omega_{\text{eff}}+f_{\text{mod}},\ 
f_5\approx0.2072=f_{\text{mod}}
.
\)
\item[(2.3)] $\omega_z=3.2$:
\(f_1=\Omega_{\text{eff}}\approx\pi/2,\ f_2\approx1.3600=\Omega_{\text{eff}}-f_{\text{mod}},\ f_3\approx1.7818=\Omega_{\text{eff}}+f_{\text{mod}},\ f_4\approx2.9311=2\Omega_{\text{eff}}-f_{\text{mod}},\ f_5\approx0.2108=f_{\text{mod}}.
\)
\end{itemize}

It is worth noting that, in a one-dimensional circle map, a rational ratio $\Omega_{\text{eff}}/f_{\text{kick}}=p/q$ guarantees a strictly periodic 
$q$-cycle inside the Arnold tongue. In contrast, our driven–dissipative dynamics exhibits quasi-periodic motion even inside the locked regions: the spectrum consistently contains small sidebands of the form $f=m\Omega_{\text{eff}}+nf_{\text{mod}}$. 
A plausible explanation is that our model is inherently high-dimensional and cannot, in general, be reduced to an exact one-dimensional circle map. As a result, the stroboscopic evolution in the locked regime does not possess an isolated $q$-cycle. Instead, the attractor retains an additional internal direction along which relaxation is extremely slow—an almost neutral mode. This weakly damped internal motion produces a gradual drift on the attractor, so that the trajectory does not close exactly after $q$ kicks. Consequently, the nominally locked oscillation is modulated by a slow intrinsic frequency $f_{\text{mod}}$, leading to the observed quasi-periodic response with sidebands. 
In the future, it would be interesting to clarify the distinct geometric structures underlying the unlocked and locked regimes and, in particular, to elucidate the dynamical origin of the modulation frequency $f_{\text{mod}}$.

\section{Finite $S$ results and matrix-free Arnold method}
\begin{figure}[ht]
\centering
\includegraphics[width=1.0\textwidth]{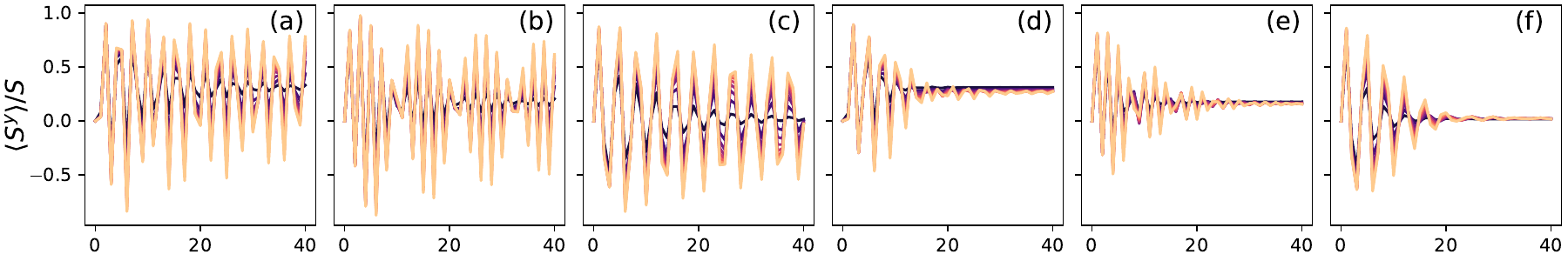}
\caption{Time evolution of $m^y=\langle \hat{S}^y\rangle/S$ for $\omega_0=1.5$, $\kappa=1.0$. The initial state is the fully polarized state with all spins up. 
(a)–(c) $\omega_z=0$; (d)–(f) $\omega_z=0.5$. 
From darkest to lightest color, $S$ increases from $10$ to $80$.  
(a), (d) $\omega_1=1.0$; (b), (e) $\omega_1=2.0$; (c), (f) $\omega_1=3.0$.}
\label{fig:few_qubit}
\end{figure}

Fig.~\ref{fig:few_qubit} presents finite-size results for $\langle \hat{S}^y\rangle/S$ in the DQTCs regime.  
The lifetime of the oscillations clearly grows with increasing $S$, and the oscillation pattern changes systematically with the kick strength, consistent with the semiclassical frequency-shift picture.    

In the main text, to obtain the Liouvillian gap, we combine exact diagonalization with a matrix-free Arnoldi Krylov approach. For exact diagonalization, we explicitly construct the Floquet--Lindbladian superoperator in a vectorized basis and diagonalize the resulting $(2S+1)^2 \times (2S+1)^2$ matrix.

As for the matrix-free Arnoldi Krylov approach, we instead employ a matrix-free representation of the one-period quantum map $\mathcal{U}$, enabling simulations at substantially larger $S$ than feasible with explicit matrix construction. Starting from a density matrix $\hat{\rho}(n)$, we integrate the Lindblad master equation over a single driving period using a fourth-order Runge--Kutta scheme and subsequently apply the kick unitary, yielding $\hat{\rho}(n+1) = \mathcal{U}(\hat{\rho}(n))$. In practice, this provides the action of $\mathcal{U}$ on a vectorized density matrix without explicitly forming the full Floquet--Lindbladian.
Beginning with a random state, we generate the Krylov subspace $\{v_1, \mathcal{U} v_1, \dots, \mathcal{U}^{m} v_1\}$ using the Arnoldi iteration with modified Gram--Schmidt orthogonalization, producing a reduced upper-Hessenberg matrix $H_{\mathrm{Hess}}$. The eigenvalues of $H_{\mathrm{Hess}}$ serve as Ritz approximations to the leading Floquet--Lindbladian eigenvalues, from which we extract the Liouvillian gap via the modulus of the first-excited eigenvalues (in our units, $\Delta = -\log|\lambda_2|$). The matrix-free Arnoldi method is benchmarked against exact diagonalization and shows excellent agreement for system sizes where both approaches are feasible.

\section{Quantum trajectory unraveling}
\begin{figure*}[htb]
\centering
\includegraphics[width=1.0\textwidth]{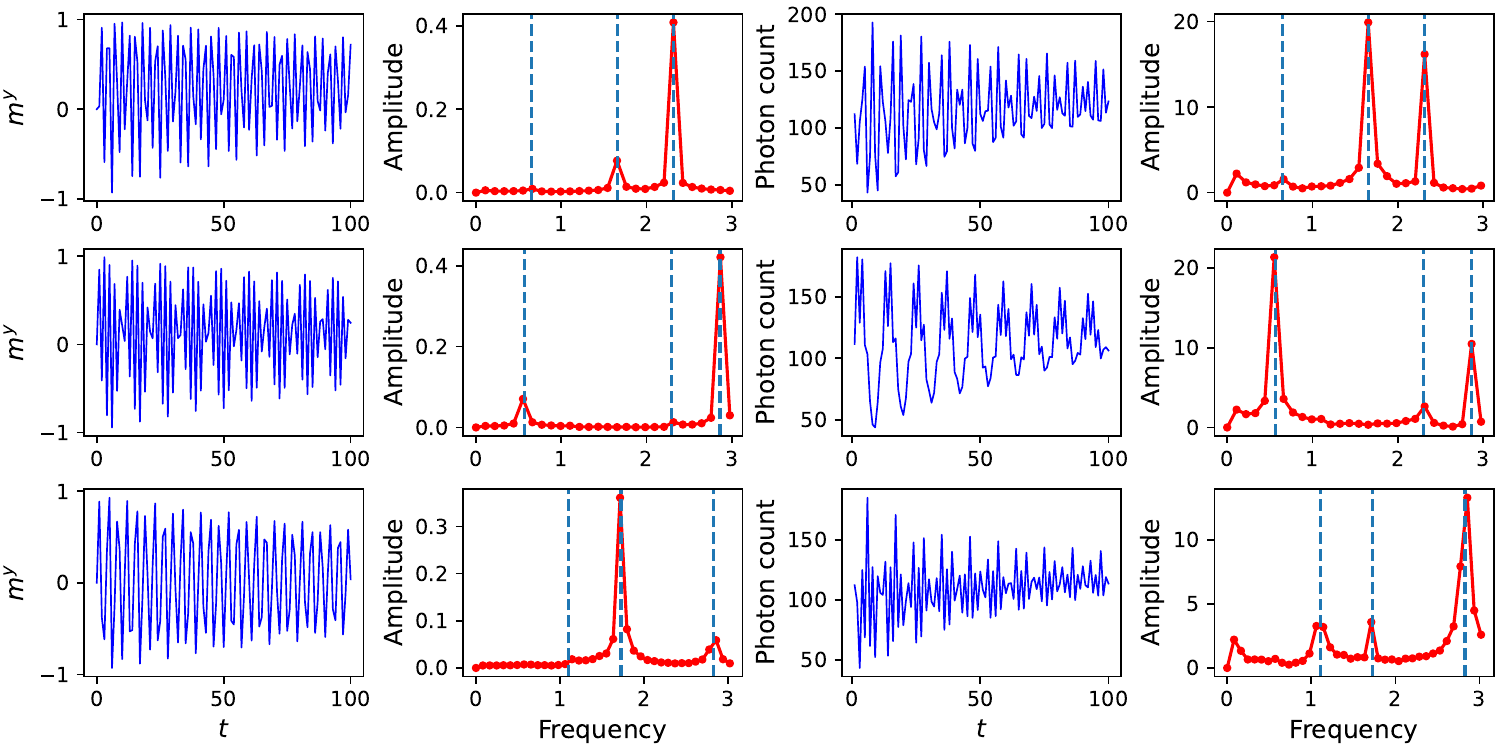}
\caption{Quantum jump trajectory for DQTCs. The presented results are averaged over 400 individual trajectories. $\omega_0=1.5$, $\omega_z=0.0$, and $S=200$. The initial state is the fully polarized state with all spins up.
Top, middle, and bottom panels correspond to $\omega_1=1.0$, $2.0$, and $3.0$, respectively. The dashed blue lines indicate the semiclassical prediction for the dominant frequencies. The observables, $m^y=\langle\hat{S}^y\rangle/S$ and the photon count, reflect the same oscillation modes and frequency structure. However, as independent dynamical quantities, the amplitudes of the dominant peaks in their Fourier spectra are distinct.
}
\label{fig:QuantumTrajectory}
\end{figure*}

To access larger $S$, we employ a quantum-trajectory (stochastic unraveling) approach, which maps the Lindblad master equation onto an ensemble of stochastic pure-state evolutions conditioned on continuous measurement outcomes.  
In the following, we focus on the quantum-jump unraveling of the Lindblad dynamics \cite{QuantumControl,MonteCarloWavefunction,QuantmTrajectoryBTC,QJQSDBTC}.

\textit{Quantum jump trajectories.—}
In a photon-counting detection scheme, one records the times at which photons emitted by the atomic system are detected.  
Conditioned on a specific detection record, the system follows a quantum jump trajectory. 

In our kicked model, the drive acts at times $t=n$ via the unitary operator $\hat{U}_K=\exp(-\mathrm{i}\hat{H}_1)$.  
Between two consecutive kicks, the dynamics are simulated using the quantum-jump method.  
We discretize time in intervals of length $\delta t$, and the conditioned state evolves under a non-Hermitian effective Hamiltonian
\begin{equation}
\hat{H}_{\text{eff}}=\hat{H}_0-\mathrm{i}\,\frac{\kappa}{2S}\hat{S}^{+}\hat{S}^{-}.
\end{equation}
At each time step $\delta t$, the state $|\Psi(t)\rangle$ either undergoes ``no-jump’’ evolution generated by $\hat{H}_{\text{eff}}$, with probability
\(
1-(\kappa/S)\langle\Psi(t)|\hat{S}^{+}\hat{S}^{-}|\Psi(t)\rangle\delta t,
\)
or it undergoes a quantum jump,
\begin{equation}
    |\Psi(t+\delta t)\rangle=\frac{\hat{S}^{-}|\Psi(t)\rangle}{\lVert\hat{S}^{-}|\Psi(t)\rangle \rVert}.
\end{equation}
Averaging over trajectories and taking the limit $\delta t\rightarrow0$ recovers the Lindblad master equation.

The trajectories shown in the first column of Fig.~\ref{fig:QuantumTrajectory} involve quantum expectation values $m^y=\langle \hat{S}^y\rangle/S$ averaged over 400 stochastic realizations.  
Such quantities are not directly accessible in a single experimental shot, due to the projective nature of quantum measurements.  
In contrast, the time record of emitted photons—shown in the third column of Fig.~\ref{fig:QuantumTrajectory}—is experimentally measurable, as demonstrated for related setups in Ref.~\cite{BTCExperiment2}. Within the quantum-jump unraveling, each detected photon corresponds to the occurrence of a jump event. 
By monitoring the photon emission statistics and performing a Fourier analysis of the photon-count signal, one can extract the intrinsic frequency of the quasi-time crystal and its tunable shift with the kick strength.

\bibliography{ref}